\documentstyle[psfig,conf_iap,]{article}
\begin{document}
\heading{%
%Begin Heading
%
Mixing of Primordial Elements in a Supernova-Driven ISM \\
%
%End Heading
} 
\par\medskip\noindent
\author{%
%Begin Author names
Miguel A. de Avillez, Mordecai-Mark Mac Low
%End Author names
}
\address{%
%First address
Department of Astrophysics, American Museum of Natural History, Central Park West at 79th Street, New York, NY 10024, USA.
}

\begin{abstract}
In order to understand the time scales for complete mixing in a
supernova-driven ISM we used the model of Avillez (2000) to follow the
mixing and dispersal of inhomogeneities with different length
scales. We included tracer fields to follow the inhomogeneities in
kpc-scale simulations with different supernova rates. We find that an
initially inhomogeneous ISM still shows local variations and a clumpy
distribution on time scales of several hundred Myr.  This clumpy
structure does not disapear over a time of a few tens of Myr, even if
the rate of SNe is increased to ten times the Galactic rate. These
local variations occur because, even in the presence of numerical
diffusion, gas does not mix quickly between hot and cold regions. The
lower limit that we are able to place on the mixing time at the kpc
scale is already longer than the time for chemistry to reach local
equilibrium in the gas. The simulations also show that the time scale
to erase inhomogeneities in the ISM is nearly independent of their
length scale over the range 25--500~pc, contrary to what is expected
from classical mixing length theory.
\end{abstract}
\section{Introduction}

Observations show significant abundance fluctuations along and between
lines of sight in the interstellar medium (ISM), suggesting the ISM is
far from homogeneous. For example, observations using the Interstellar
Medium Absorption Profile Spectrograph (IMAPS) show local variations
in the D/H ratio of about a factor of three (Jenkins et al.~1999;
Sonneborn et al.~2000). Deuterium is mainly produced during Big Bang
nucleosynthesis\footnote{However, Mullan \& Linsky (1999) have
proposed that D can be formed in stellar flares from M dwarf stars.}
and only destroyed in pre-main sequence stars. Therefore, the D in the
ISM traces the fraction of gas not processed through stars, and the
detection of local variations in the D/H ratio indicates incomplete
mixing of processed material into the ISM. However, such an
inhomogeneity picture contradicts the almost homogeneous distribution
seen in O/H, which suggest that the ISM mixes completely on timescales
no longer than that for chemistry reach an equilibrium.

These observations pose the question of how chemical elements such as
D can avoid being mixed well enough to suppress local variations in
the D/H ratio. Furthermore, the ISM is regulated by supernova
explosions that form well structured and explosive flows in addition
to diffuse turbulence that acts on the smaller scales. It is
therefore, necessary to understand the mixing process in such a medium
by carrying out direct simulations of the evolution of the ISM
polluted with inhomogeneities of different length scales.

Section 2 deals with the description of the three-dimensional model
used in this study, as well as, with the setup and evolution of the
simulations. In section 3 we present the results of these simulations
and discuss them in the light of the classical diffusion
theory. Finally, in section 4 we present a summary of the main results.
\begin{table}
\centering
\caption{Simulation parameters for the runs described in this paper.}
\begin{tabular}{cllcc}
\hline
Run$^a$ & $\Delta x^{b}$ & $\sigma/\sigma_{\rm Gal}^{c}$ & $\tau_{50}^{d}$ & $\tau_{500}^{d}$ \\
        &    [pc]        &      &  [Myr]  &  [Myr] \\
\hline
L11  & 5    & 1  & 303.1 & 311.2 \\
L12  & 5    & 10 & 143.0 & 167.2 \\
L13  & 5    & 20 & 123.5 & 144.9 \\
L16  & 5    & 30 & 108.4 & 130.5 \\
L17  & 5    & 50 & 107.0 & 127.2 \\
L21  & 2.5  & 1  & 318.8 & 326.5 \\
L22  & 2.5  & 10 & 162.4 & 184.5 \\
L23  & 2.5  & 20 & 138.5 & 151.6 \\
L31  & 1.25 & 1  & 347.2 & 366.4 \\
L32  & 1.25 & 10 & 180.1 & 199.1 \\
L33  & 1.25 & 20 & 149.6 & 163.8 \\
\hline
\multicolumn{5}{l}{$^a$ Run number and finest grid level number.}\\
\multicolumn{5}{l}{$^b$ Finest grid resolution.}\\
\multicolumn{5}{l}{$^c$ Time between SNe in terms of the Galactic value.}\\
\multicolumn{5}{l}{$^d$ Time for complete mixing.} \\
\end{tabular}
\end{table}
\section{Supernova-Driven ISM Model}
\subsection{Mixing Simulations}
The model includes a fixed gravitational field provided by the stars
in the disk, and an ideal-gas equation of state.  Radiative cooling is
treated assuming collisional ionization equilibrium, using a piecewise
cooling function following that presented in Figure~2 of Dalgarno \&
McCray (1972) with an ionization fraction of 0.1 at temperatures below
$10^4$~K and a temperature cutoff at 10~K. Background heating due to
starlight varies with $z$ as described in Wolfire et al. (1995), and
at $z=0$ it is chosen to balance radiative cooling at 8000 K.

Type Ib, Ic, and II~SNe are set up at the beginning of their Sedov
phases, with radii determined by their progenitor masses, as described
in detail by Avillez (2000). Sixty percent of the SNe are set up
within associations, and the rest are set up at random sites. OB
associations are set up in a layer with a scale height of 46 pc (from
the midplane) following the distribution of the molecular gas in the
Galaxy, while the isolated SNe are set with an exponential
distribution with a scale height of 90 pc.  The rates of occurrence of
SNe in the Galaxy are normalized to the volume under study, with the
Galactic rate for types Ib and Ic in the Galaxy taken to be $2\times
10^{-3}$ yr$^{-1}$, while for type~II it is $1.2\times 10^{-2}$
yr$^{-1}$ (Cappellaro et al. 1997). The total rate of these SNe in the
Galaxy is $\tau_{\rm gal} = 1.4\times 10^{-2}$ yr$^{-1}$, or one SN
every 71 yr.

The interstellar gas is initially distributed in a smooth disk with
the vertical distribution of Dickey \& Lockman (1990). In addition, an
exponential profile representing the $z-$distribution of the warm
ionized gas with a scale-height of 1 kpc in the Galaxy as described in
Reynolds (1987) is used.

We model inhomogeneities in the ISM with a tracer field distributed in
a checkerboard pattern of empty and full squares. We followed the
evolution of models with squares having length $L$ of 25, 50 and 500
pc in the $x$ and $y-$directions (parallel to the Galactic
plane). Slab symmetry of the tracer field is assumed along the
$z$-direction (perpendicular to the Galactic plane). In the current
work we report on simulations using five SN rates: $\sigma /
\sigma_{\rm gal} =1$, 10, 20, 30 and 50. The simulations are run until
400 Myr for $\sigma /\sigma_{\rm gal} =1$ and 200 Myr for the
remaining supernova rates. This evolution time is enough for the
system to reach steady state. A summary of these runs is presented in
Table 1.

The computational domain contains a section of the Galaxy with an area
of 1 kpc$^{2}$ and vertical extension from -10 to 10 kpc.  The
innermost edge lies 8.5 kpc from the Galactic centre.  The
computational grid has a resolution of 10 pc, except in the layer
between -250 and 250 pc, where three levels of adaptive mesh
refinement are used, yielding a finest resolution of 1.25 pc. These
simulations use the piecewise-parabolic method of Colella \& Woodward
(1984), a third-order scheme implemented in a dimensionally-split
(Strange 1968) manner in combination with the AMR scheme of Berger \&
Colella (1989), with the subgridding scheme of Bell et al.\ (1994).

\subsection{Global Evolution of the ISM}
Once disrupted by the SN explosions, which start occurring at time
$t=0$, the disk never returns to its initial state. Instead,
regardless of the initial vertical distribution of the disk gas, a
thin disk of cold gas forms in the Galactic plane, and, above and
below, a thick, inhomogeneous gas disk forms.  The code does not
explicitly follow ionization states, but we can trace gas with
temperature $T\leq 10^{4}$ K and scale height of 180~pc, which we
designate H{\sc i}, and gas with $10^{4} \le T\le 10^{5}$ K and scale
height of 1~kpc, which we designate H{\sc ii}. These distributions
reproduce those described in Dickey \& Lockman (1990) and Reynolds
(1987), respectively.  The upper parts of the thick H{\sc ii} disk
form the disk-halo interface, where a large scale fountain is set up
by hot ionized gas escaping in a turbulent convective flow (see Avilez
2000 and Avillez \& Mac Low 2001).
\begin{figure}
\centering{\hspace*{-0.1cm}}
\psfig{file=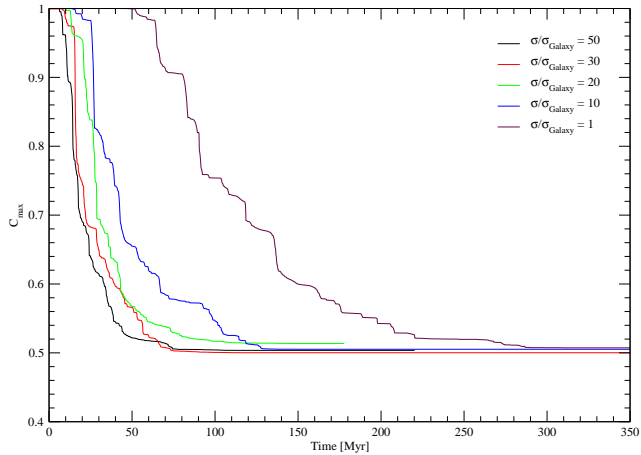,angle=-90,width=0.8\hsize,clip=}
\caption{Time evolution of the maximum $C_{\rm max}$ of the tracer field for inhomogeneities with length scales of 50 pc, for $\sigma / \sigma_{\rm gal} =1$, 10, 20, 30 and 50 and a fine grid resolution of 5 pc.}
\label{mavillez_fig1}
\end{figure}

\section{Evolution of ISM Inhomogeneities}

The tracer field is a scalar whose mass flux follows that of the
density and whose value is unity in the regions of the flow where it
is present and zero elsewhere in the flow. By definition the tracer
field should keep its value of 1, and therefore its average value
$<C>$ on a checkerboard should always be 0.5 regardless of how it is
advected. Numerical schemes introduce numerical diffusion leading
sooner or later to a change of the value of the tracer field. However,
its average value over the grid should remain constant. The
simulations show that as the disk gas evolves to steady state $<C>$
approaches 0.5, with variations of no more than 2.4\% occurring for a
supernova rate of 30 times that in the Galaxy.
\begin{figure}
\centering{\hspace*{-0.1cm}}
\psfig{file=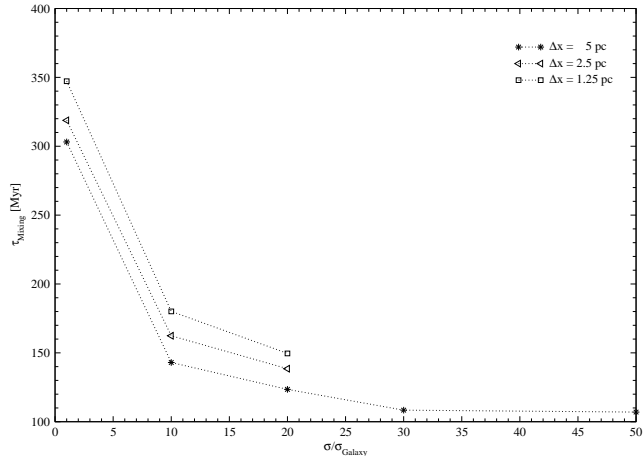,angle=-90,width=0.8\hsize,clip=}
\caption{Variation of the mixing time with supernova rate for finer grid resolutions $\Delta x=1.25,~2.5,~5$ pc. Mixing time increases by factor between 1.05 and 1.2 when resolution is incresed by a fctor of 2.}
\label{mavillez_fig2}
\end{figure}
We define complete mixing to be when the maximum and minimum of the
tracer field become numerically indistinguishable from each other and
$<C>$, i.e., $C_{\rm max}=C_{\rm min}=<C>\sim0.5$ ($\Delta C=C_{\rm
max}-C_{\rm min} < 0.01$). Table 1 presents the time scales for
complete mixing in checkerboards with 50 and 500 pc scale-lengths for
resolutions of 5, 2.5 and 1.25 pc, while Figures 1 and 2 show the time
variation of $C_{\rm max}$ for a resolution of 5 pc and the variation of the
mixing time with supernova rate for different finer resolutions, respectively.

Figure 1 shows that there is a steady decrease of $C_{\rm max}$ until
complete mixing occurs. Futhermore, $C_{\rm max}$ and $C_{\rm min}$
take almost the same amount of time to vary from a difference $\Delta
C=1$ to $0.05$ as the time from $\Delta C=0.05$ to $0.01$, when we
define complete mixing to occur. This indicates that smaller scale
structures dominate the mixing process after the gas has been
completely mixed on larger scales.

Figures 1 and 2 show that the time taken for complete mixing to occur
decreases with increasing supernova rate, that is, increased energy
input into the ISM. However, the rate of increase of mixing reaches a
saturation level for $\sigma/\sigma_{Galaxy}> 20$, beyond which the
increase in the supernova rate does not accelerate mixing in the
ISM. For SN rates greater than 20 times the Galactic rate we see that
the mixing time scales differ by only a few Myr (see Table 1), with
complete mixing occurring at $\sim107$ Myr for $\Delta x=5$ pc. This
results from the fact that for $\sigma/\sigma_{Galaxy}\ge20$, after
the first 50 Myr of evolution a large fraction fo the disk gas has a
temperature greater than $10^{5}$ K and at 100 Myr of evolution all
the gas is hotter than $10^{5}$ K. As a consequence of the high SN
rate the disk becomes hotter and cooling becomes inefficient.  At this
high temperature the sound speed is correspondingly high, so the gas
mixes within a 40-50 Myr period (Roy \& Kunth 1995) leading to
complete mixing at some 110 Myr.

The time taken for complete mixing to occur in a checkerboard with 50
pc squares varies from $\sim300$ Myr for $\sigma/\sigma_{Galaxy}=1$ to
some 120 Myr for $\sigma/\sigma_{Galaxy}= 20$ and 107 for $\sigma/\sigma_{Galaxy}>30$ for $\Delta x=5$
pc, while for $\Delta x=1.25$ pc mixing varies from 347 Myr for
$\sigma/\sigma_{Galaxy}=1$ to 150 Myr for
$\sigma/\sigma_{Galaxy}=20$. On 500 pc squares complete mixing is
surprisingly only delayed by around 20 Myr (Table 1). The similarity
between these time scales indicates that the mixing process is almost
independent of the length-scale of the inhomogeneities, and therefore,
of the initial setup adopted for the tracer field.  Rather, it is
regulated by supernovae and their energy input into the ISM.

According to classical mixing length theory the mixing time scales of
these boards should follow the relation $\tau \propto \ell^2$, where
$\ell$ is the length of the squares. Therefore, the mixing time of
inhomogeneities with a scale length of 500 pc should be 25 times
longer than the mixing time for 50 pc squares.  The computed
similarity in the mixing time scales (Table 1) for the different
inhomogeneity sizes shows that classical mixing length theory does not
describe mixing in the ISM well.  This seems to be because much of the
ISM is dominated by nearly laminar flows inside of SNRs and
superbubbles rather than by diffuse turbulence.  The latter does act
on smaller scales and in cooler gas, in which classical mixing length
theory may apply better.

\section{Summary}

In this paper we try to shed light on the time scales that it takes to
erase inhomogeneities in the ISM and how this process depends on the
inhomogeneity length scale.  We find: (1) The time scale to erase
inhomogeneities in the ISM is nearly independent of their length scale
on scales above 10 pc or so; (2) Mixing time scales decrease with
increasing supernova rate until all the gas is hot and the mixing rate
saturates.  This occurs when $\sigma/\sigma_{\mbox{Galaxy}}\ge 20$,
when the mixing time scale is $\sim 107-120$ Myr for $\Delta x=5$ pc and
150 Myr for $\Delta x=1.25$ pc; (3) Even after some $80\%$ of the gas
has been mixed, the simulations show that the ISM remains clumpy for
at least twice as long if we examine scales smaller than a kpc. Even
if the rate of SNe is increased to ten times the Galactic rate, this
clumpy structure does not disappear over a time of a few tens of Myr.

This work shows that inhomogeneities that are present in the ISM,
assuming that no further inhomogeneities are introduced into the
system, take a very long time to be erased. The time for complete
mixing (greater than 107 Myr, which occurs for
$\sigma/\sigma_{\mbox{Galaxy}}= 50$ and $\Delta x=5$ pc) is already
longer that any time scale for chemical equilibrium to occur, and
therefore, the ISM will show a clumpy distribution. The local
variations occur because, even in the presence of numerical diffusion,
gas does not mix quickly between hot and cold regions.

\acknowledgements{
The authors acknowledge useful discussions with D. York, and
E. Jenkins, and with R. Ferlet, who also emphasized the importance of
working on this problem. This work was supported by an NSF CAREER
grant (AST 99-85392).}

\vfill
\end{document}